\documentclass[aps,preprint,showpacs,nofootinbib,showkeys,tightenlines]{revtex4}
\usepackage{graphicx}  %
\usepackage{amsmath}
\usepackage{bm}  %

\newcommand{\bea}{\begin{eqnarray}}
\newcommand{\eea}{\end{eqnarray}}
\newcommand{\beq}{\begin{equation}}
\newcommand{\eeq}{\end{equation}}
\newcommand{\bqa}{\begin{eqnarray}}
\newcommand{\eqa}{\end{eqnarray}}

\def\mqo2{{\!\!\!}}
\renewcommand{\Im}{{\rm Im\,}}

\begin{document}

\title{
Universal Relation for the Inelastic Two-Body Loss Rate}
\author{Eric Braaten}
%\email{braaten@mps.ohio-state.edu}
\affiliation{Department of Physics,
         The Ohio State University, Columbus, OH\ 43210, USA\\}
\affiliation{Helmholtz-Institut f\"ur Strahlen- und Kernphysik
	and Bethe Center for Theoretical Physics,
	Universit\"at Bonn, 53115 Bonn, Germany\\}
\author{H.-W. Hammer}
\affiliation{Helmholtz-Institut f\"ur Strahlen- und Kernphysik
	and Bethe Center for Theoretical Physics,
	Universit\"at Bonn, 53115 Bonn, Germany\\}
\affiliation{Institut f\"ur Kernphysik, 
Technische Universit\"at Darmstadt, 64289 Darmstadt, Germany\\}
\affiliation{ExtreMe Matter Institute EMMI, GSI Helmholtzzentrum 
f\"ur Schwerionenforschung GmbH, 64291 Darmstadt, Germany\\}
\date{\today}
%\date{November 2007}

\begin{abstract}
Strongly-interacting systems consisting of 
particles that interact through a large scattering length
satisfy universal relations that relate many of their central 
properties to the contact, which measures 
the number of pairs with small separations.
We use the operator product expansion of quantum field theory to
derive the universal relation for the inelastic 2-body loss rate.
A simple universal relation between the loss rate and the contact is 
obtained by truncating the expansion after the lowest dimension operator.
We verify the universal relation explicitly by direct calculations  
in the low-density limit at nonzero temperature.
This universal relation can be tested experimentally using
ultracold quantum gases of atoms in hyperfine states
that have an inelastic spin-relaxation channel.
\end{abstract}

\smallskip
\maketitle

%%%%%%%%%%%%%%%%%%%%%%%%%%%%%%%%%%%%%%%%%%%%%%%
\section{Introduction}
\label{sec:intro}
%%%%%%%%%%%%%%%%%%%%%%%%%%%%%%%%%%%%%%%%%%%%%%%

Atoms whose scattering length 
is large compared to the range of their interactions have universal 
properties that depend only on the large scattering length and
are insensitive to details of their interactions at short distances \cite{Braaten:2004rn}.
These properties are universal in the sense that they apply to other types of 
nonrelativistic particles with short-range interactions, such as hadrons and nuclei,
provided those particles have the same symmetries and mass ratios 
and large scattering lengths.
The strong correlations associated with the large scattering length 
present challenges to understanding and calculating the 
properties of many-body systems of such particles.  
These challenges are particularly acute in the unitary limit 
in which the scattering length is infinite.  This limit is characterized by 
scale invariance, so there are no well-defined quasiparticles 
upon which to base our understanding of the system.

Relations between various properties 
of a system that depend only on the large scattering length
are called {\it universal relations}.  Such relations provide 
powerful constraints on the behavior of the system 
that must be respected in spite of the strong correlations.
Shina Tan derived a number of universal relations for 
the strongly-interacting two-component Fermi gas,
which consists of two types
of fermions that interact through a large scattering 
length \cite{Tan0505,Tan0508,Tan0803}.
These relations apply to any state of the system:
few-body or many-body, homogeneous or trapped,
normal or superfluid, balanced or polarized.  
For a few-body system, 
they apply as long as the energies relative to the appropriate threshold are small 
compared to the energy scale set by the range of the interactions.
For a many-body system, 
they apply as long as the temperature and number density 
are small compared to the energy and density scales set by the range.

Tan's universal relations all involve a property of the system 
$C$ called the {\it contact}
that measures the number of pairs of atoms of the two types with
small separations.  The contact is an extensive quantity 
that can be expressed as the integral over space of the contact density:  
$C = \int d^3r \, {\cal C}$.  
A convenient operational definition of the contact is provided by Tan's 
{\it adiabatic relation}, which gives the rate of change of the total energy $E
$ of the system as the scattering length is varied~\cite{Tan0508}:
%-----------------
\begin{equation}
\frac {d \ \ }{da^{-1}} E = - \frac{\hbar^2}{8 \pi \mu} ~ C ,
\label{adiabatic-Tan}
\end{equation}
%-----------------
where $\mu$ is the reduced mass of the two types of atoms.
If the system is at nonzero temperature,
the derivative in Eq.~(\ref{adiabatic-Tan}) should be taken 
with the entropy held fixed.
Using the adiabatic relation, 
any theoretical method that can be used to calculate the total energy 
of the system can also be used to calculate the contact.
If the total energy can be measured in an experiment
and if the scattering length can be controlled experimentally,
for example by using a Feshbach resonance,
the adiabatic relation can be used to measure the contact \cite{Jin1002}.

Tan derived his universal relations from the many-body Schr\"odinger 
equation by using generalized functions designed to implement the 
Bethe-Peierls boundary conditions associated with the large scattering 
length~\cite{Tan0505,Tan0508,Tan0803}.  
They were rederived by Braaten and Platter 
using the machinery of quantum field theory, specifically renormalization 
and the operator product expansion \cite{Braaten:2008uh}.
Tan's  universal relations make it clear that the contact 
plays a central role in the strongly-interacting Fermi gas.
Subsequent work 
revealed that several of the most important experimental probes of 
ultracold atoms are related to the contact,
including radio-frequency spectroscopy~\cite{PZ0707,BPYZ0707} 
and photoassociation~\cite{WTC0807,Zhang-Leggett}.
A review of universal relations for fermions with a large scattering length 
was presented in Ref.~\cite{Braaten:2010if}.

One of the universal relations given in Ref.~\cite{Braaten:2010if} 
is for the inelastic 2-body loss rate for fermions with a large scattering length.
If atoms of types 1 and 2 have inelastic scattering channels 
with a large energy release, low-energy atoms are lost 
from the system through inelastic collisions.
Because of the inelastic scattering channels, 
the scattering length $a$ must be complex with a negative imaginary part. 
The rate at which the number of either type of atom decreases is
proportional to the contact:%
\footnote{The relation actually stated in Ref.~\cite{Braaten:2010if}
is that the rate of change of the number density $n_\sigma$
is proportional to the contact density ${\cal C}$, which is not strictly correct.
Integration over space is necessary to eliminate a term
involving the divergence of the probability current.
}
%-----------------
\begin{equation}
\frac {d \ }{dt} N_\sigma = - \frac{\hbar}{4 \pi \mu} ~ {\rm Im}(1/a) ~ C .
\label{lossrate}
\end{equation}
%-----------------
The fact that the loss rate is proportional to $C$ was first pointed 
out by Tan~\cite{Tan-private}.
The prefactor of $C$ in Eq.~(\ref{lossrate}) was first written 
down by Braaten and Platter~\cite{Braaten:2008uh}.  
To the best of our knowledge, a derivation of 
Eq.~(\ref{lossrate}) has not been presented previously. 
In this paper, we derive the universal relation
for the 2-body loss rate using the operator product expansion of 
quantum field theory.
The simple universal relation in Eq.~(\ref{lossrate})  is 
obtained by truncating the expansion after the lowest dimension operator.

%%%%%%%%%%%%%%%%%%%%%%%%%%%%%%%%%%%%%%%%%%%%%%%%%%%%%%%%%%%%%%%%%%%%%%%%%%
\section{Inelastic two-body loss}
\label{sec:lossrate}
%%%%%%%%%%%%%%%%%%%%%%%%%%%%%%%%%%%%%%%%%%%%%%%%%%%%%%%%%%%%%%%%%%%%%%%%%%

Inelastic two-atom scattering channels with a large energy release
cause the loss of low-energy atoms 
from the system.  
The inelastic scattering channel may consist of lower hyperfine states
of the atoms of types 1 and 2.
The loss of low-energy atoms is actually due to 
their transformation into high-energy atoms.  
If the system is in a trapping potential, the high-energy atoms
may have enough energy to escape from the trap.   Even if they do not escape,
their effects on the low-energy atoms can be neglected 
if their interactions with the low-energy atoms are weak enough.
If the energy gap $E_{\rm gap}$ between the threshold for atoms 1 and 2 
and the threshold for the final-state atoms 
is much larger than the low energies of interest, 
the inclusive effects of the inelastic collisions can be taken into account through
anti-Hermitian terms in the Hamiltonian.
These terms must be anti-Hermitian to account for the loss of
probability from transitions of pairs of low-energy atoms to pairs of high-energy atoms, 
which are not taken into account explicitly.

An important consequence of the large energy gap $E_{\rm gap}$ 
is that there must be a small separation between the atoms of types 1 and 2 that disappear.
By conservation of energy, the high-energy atoms 
created by the inelastic scattering of atoms of types 1 and 2
will emerge with large relative momentum that is approximately $(\mu E_{\rm gap})^{1/2}$.
By the uncertainty principle, the localized wave packet 
associated with each outgoing atom can be traced back to a region 
from which the atom emerged whose size is approximately 
$\hbar/(\mu E_{\rm gap})^{1/2}$.  But the two regions 
from which the two final-state atoms emerge
must also be separated by a distance
at most of order $\hbar/(\mu E_{\rm gap})^{1/2}$,
because the interactions that change the spin states of the atoms
must also deliver a momentum kick of order $(\mu E_{\rm gap})^{1/2}$.
This large momentum can be transferred to the two 
final-state atoms only if they are close together.
Even if the momentum is transferred through an intermediate
atom in the original state 1 or 2, that atom must be far off its energy shell 
and therefore can propagate only over a short distance.
The two atoms that disappear must therefore have a separation 
of order $(\mu E_{\rm gap})^{1/2}$.
Thus the non-Hermitian terms in the Hamiltonian must have short range.

%%%%%%%%%%%%%%%%%%%%%%%%%%%%%%%%%%%%%%%%%%%%%%%
\section{Derivation of the Universal Relation}
\label{sec:Derivation}
%%%%%%%%%%%%%%%%%%%%%%%%%%%%%%%%%%%%%%%%%%%%%%%

In this section, we derive the universal relation in Eq.~(\ref{lossrate})
between the inelastic 2-atom loss rate and the contact.
We use quantum field theory 
methods similar to those used in Ref.~\cite{Braaten:2008uh} 
to give concise derivations of other universal relations.

%%%%%%%%%%%%%%%%%%%%%%%%%%%%%%%%%%%%%%%%%%%%%%%%
\subsection{Quantum field theory}
\label{sec:QFT}
%%%%%%%%%%%%%%%%%%%%%%%%%%%%%%%%%%%%%%%%%%%%%%%%

We begin by presenting the quantum field theory formulation of the 
problem of atoms with a large real scattering length $a$
in the zero-range limit.
We introduce quantum field operators $\psi_1(\bm{r})$ and $\psi_2(\bm{r})$ 
that annihilate atoms of type 1 and 2, respectively.  
If either atom is a boson, the fields satisfy the canonical commutation
relations.  If they are both fermions, they satisfy anticommutation
relations.  The masses of the atoms are $m_1$ and $m_2$, 
and their reduced mass is $\mu = m_1 m_2/(m_1 + m_2)$.
The Hamiltonian operator for the system of atoms is the sum of a kinetic term, 
an interaction term, and possibly a term describing an external trapping potential: 
$\hat H = \hat T + \hat U + \hat V$.
The kinetic and interaction terms can be expressed as
%-----------------
\begin{subequations}
\begin{eqnarray}
\hat T &=& 
\sum_\sigma \frac{1}{2 m_\sigma} \int \!\! d^3r \, 
\bm{\nabla} \psi_\sigma^\dagger \cdot \bm{\nabla} \psi_\sigma (\bm{r}),
\label{T}
\\
\hat U &=& 
\frac{\hbar^2 g_0}{2 \mu} \int \!\! d^3r \, 
\psi_1^\dagger \psi_2^\dagger \psi_2 \psi_1 (\bm{r}),
\label{U}
\end{eqnarray}
\end{subequations}
%-----------------
where $g_0$ is the bare coupling constant, which depends on the 
ultraviolet momentum cutoff $\Lambda$.  The products of quantum
field operators followed by the argument $(\bm{r})$ indicate 
that the operators are all evaluated at the same position $\bm{r}$.
In the fermionic case, the ordering of the fields in the interaction term
guarantees the positivity of the local operator. 
The ultraviolet cutoff can be implemented by limiting the Fourier 
transforms of the quantum field operators to wavenumbers satisfying
$|{\bm k}| < \Lambda$.  The scattering length for a pair of atoms 
of types 1 and 2 will have the value $a$ if the bare
coupling constant is chosen to be
%-----------------
\begin{equation}
g_0 = \frac{4 \pi}{1/a - 2 \Lambda/\pi} .
\label{g}
\end{equation}
%-----------------

We now consider the problem of atoms that have inelastic scattering channels
with a large energy gap.  The large energy gap ensures that the 
anti-Hermitian terms in the Hamiltonian density must have a short range.
We can therefore take them to be local operators involving the quantum fields 
$\psi_1$ and $\psi_2$.
Since the two atoms that disappear must have a very small separation,
the appropriate operator is $\psi_1^\dagger \psi_2^\dagger \psi_2 \psi_1$,
which annihilates a pair of particles at a point and then creates them again.
This is precisely the operator in the interaction term 
of the Hamiltonian in Eq.~(\ref{U}).
To take account the loss of atoms, we can add such a term to the Hamiltonian 
with a coefficient that is pure imaginary and negative.  
Equivalently, we can simply take the bare coupling constant $g_0$ in 
Eq.~(\ref{U}) to have a negative imaginary part.  
The inelastic scattering channel gives
the scattering length $a$ a negative imaginary part.
The complex bare coupling constant $g_0$ 
can be obtained simply by inserting that complex scattering length $a$
into Eq.~(\ref{g}).

%%%%%%%%%%%%%%%%%%%%%%%%%%%%%%%%%%%%%%%%%%%%%%%
\subsection{Particle loss rate}
\label{sec:plossrate}
%%%%%%%%%%%%%%%%%%%%%%%%%%%%%%%%%%%%%%%%%%%%%%%%

We now consider the loss rate of the atoms.
The time evolution of a state $| X(t) \rangle$ in the 
quantum field theory is given by
%-----------------
\begin{equation}
| X(t) \rangle = \exp(-i \hat H t/\hbar) | X(0) \rangle.
\label{X-t}
\end{equation}
%-----------------
Because $\hat H$ has a non-Hermitian part,
the norm of the state $| X(t) \rangle$ decreases with time.
The average number density of atoms of type $\sigma$ 
in the state $| X(t) \rangle$ is given by the expectation value 
of the  number density operator:
%-----------------
\begin{equation}
n_\sigma(\bm{r},t) = 
\frac{\langle X(t) | \psi_\sigma^\dagger \psi_\sigma (\bm{r})| X(t) \rangle}
{\langle X(t) | X(t) \rangle}
\equiv
\big\langle \psi_\sigma^\dagger \psi_\sigma (\bm{r}) \big\rangle .
\label{n-t}
\end{equation}
%-----------------
Inserting the time evolution from Eq.~(\ref{X-t})
and differentiating with respect to time, we get
%-----------------
\begin{equation}
\frac{d \ }{dt} n_\sigma(\bm{r},t) = \frac{i}{\hbar}
\left( \big\langle \hat H^\dagger \, \psi_\sigma^\dagger \psi_\sigma (\bm{r}) 
	- \psi_\sigma^\dagger \psi_\sigma (\bm{r}) \, \hat H \big\rangle 
- \big\langle \psi_\sigma^\dagger \psi_\sigma (\bm{r}) \big\rangle
\big\langle \hat H^\dagger - \hat H \big\rangle \right) .
\label{dndt}
\end{equation}
%-----------------
The contribution from the kinetic term in the Hamiltonian 
can be expressed as the divergence of a current:
%-----------------
\begin{equation}
\big\langle \big[ \hat T, 
        \psi_\sigma^\dagger \psi_\sigma (\bm{r}) \big] \big\rangle =
 i \hbar \bm{\nabla} \cdot\
\big\langle \bm{J}_\sigma (\bm{r}) \big\rangle,
\label{dndt-T}
\end{equation}
%-----------------
where $\bm{J}_\sigma$ is the probability current operator 
for atoms of type $\sigma$:
%-----------------
\begin{equation}
\bm{J}_\sigma(\bm{r}) =
- \frac{i \hbar}{2 m_\sigma}
\left( \psi_\sigma^\dagger \bm{\nabla} \psi_\sigma (\bm{r}) 
- \bm{\nabla} \psi_\sigma^\dagger \psi_\sigma (\bm{r})\right).
\label{J-sigma}
\end{equation}
%-----------------
The flow term in Eq.~(\ref{dndt-T})
can be eliminated by integrating over all space 
and using the divergence theorem.  The resulting expression 
for the rate of change of the total number of particles of type $\sigma$ is
%-----------------
\begin{equation}
\frac{d \ }{dt} N_\sigma = 
\frac{i}{\hbar} \int \!\! d^3 r \, 
\left( \big\langle \hat U^\dagger \, \psi_\sigma^\dagger \psi_\sigma (\bm{r})
- \psi_\sigma^\dagger \psi_\sigma (\bm{r}) \, \hat U \big\rangle
- \big\langle \psi_\sigma^\dagger \psi_\sigma (\bm{r}) \big\rangle 
\big\langle \hat U^\dagger - \hat U \big\rangle \right) ,
\label{dNdt}
\end{equation}
%-----------------
where $\hat U$ is the interaction energy operator in Eq.~(\ref{U}).
Using the (anti)commutation relations for the field operators, 
$\hat U$ can be expressed 
in terms of the number density operators $\psi_1^\dagger \psi_1$
and $\psi_2^\dagger \psi_2$.  It therefore commutes with
$\psi_\sigma^\dagger \psi_\sigma (\bm{r})$.  Thus only the non-Hermitian part 
of $\hat U$ contributes to Eq.~(\ref{dNdt}).
The prefactor in $\hat U^\dagger - \hat U$ can be expressed as
%-----------------
\begin{equation}
\frac{\hbar^2 (g_0^* - g_0)}{2 \mu} =
i \, {\rm Im}(1/a) \frac{\hbar^2 |g_0|^2}{4 \pi \mu}.
\label{Img}
\end{equation}
%-----------------
Thus the loss rate in Eq.~(\ref{dNdt}) reduces to
%-----------------
\begin{eqnarray}
\frac{d \ }{dt} N_\sigma &=&
- \frac{\hbar \, {\rm Im}(1/a)}{4 \pi \mu} 
\int \!\! d^3 r \int \!\! d^3 r' \left(
\big\langle |g_0|^2 \psi_1^\dagger \psi_2^\dagger \psi_2 \psi_1 (\bm{r}') \,
\psi_\sigma^\dagger \psi_\sigma (\bm{r}) \big\rangle \right.
\nonumber
\\
&& \left. \hspace{5cm}
- \big\langle |g_0|^2 \psi_1^\dagger \psi_2^\dagger \psi_2 \psi_1 (\bm{r}')  \big\rangle 
\big\langle \psi_\sigma^\dagger \psi_\sigma (\bm{r}) \big\rangle \right) .
\label{dNdt-simple}
\end{eqnarray}
%-----------------
We have combined the factor of $|g_0|^2$ with the operator
$\psi_1^\dagger \psi_2^\dagger \psi_2 \psi_1$, because the dependence of 
matrix elements of this operator on the ultraviolet cutoff $\Lambda$
is exactly cancelled by the $\Lambda$-dependence of $|g_0|^2$ \cite{Braaten:2008uh}.
The loss rate has been expressed in terms of a double integral over space 
of the correlator of the contact density operator and the number density operator.
The subtraction in Eq.~(\ref{dNdt-simple}) removes
the disconnected part of the correlator.
Thus the loss rate can be expressed more concisely as
%-----------------
\begin{eqnarray}
\frac{d \ }{dt} N_\sigma &=&
- \frac{\hbar \, {\rm Im}(1/a)}{4 \pi \mu} 
\int \!\! d^3 r \int \!\! d^3 r' 
\big\langle |g_0|^2 \psi_1^\dagger \psi_2^\dagger \psi_2 \psi_1 (\bm{r}') \,
\psi_\sigma^\dagger \psi_\sigma (\bm{r}) \big\rangle_{\rm connected},
\label{dNdt-concise}
\end{eqnarray}
%-----------------
where the subscript ``connected'' indicates the connected part of the 
matrix element.

%%%%%%%%%%%%%%%%%%%%%%%%%%%%%%%%%%%%%%%%%%%%%%%
\subsection{Operator product expansion}
\label{sec:OPE}
%%%%%%%%%%%%%%%%%%%%%%%%%%%%%%%%%%%%%%%%%%%%%%%

We proceed to use the operator product expansion (OPE) 
to reduce the expression for the loss rate to single integrals over space 
of expectation values of local operators.
If the separation $|\bm{r}'-\bm{r}|$ of the operators 
in the first term in the integrand in Eq.~(\ref{dNdt-simple}) is much larger than
the correlation length of the system, the expectation value of the 
operator product factors into the product of the expectation values
and it is therefore cancelled by the second term.
Thus the integrand goes to zero as  $|\bm{r}'-\bm{r}| \to \infty$.
This motivates the expansion of the operator product  in powers of $|\bm{r}'-\bm{r}|$:
%-----------------
\begin{eqnarray}
|g_0|^2 \psi_1^\dagger \psi_2^\dagger \psi_2 \psi_1 (\bm{r}') \,
\psi_\sigma^\dagger \psi_\sigma (\bm{r})
= \delta^3(\bm{r}'-\bm{r}) \, |g_0|^2 \psi_1^\dagger \psi_2^\dagger \psi_2 \psi_1 ((\bm{r} + \bm{r}')/2)
\nonumber \\
+ \sum_n f_{\sigma n}(\bm{r}'-\bm{r}) \, {\cal O}_n ((\bm{r} + \bm{r}')/2).
\label{OPE}
\end{eqnarray}
%-----------------
This operator equation is the {\it operator product expansion}.
The sum is over local operators ${\cal O}_n$ with increasingly higher scaling dimensions.
The coefficients $f_{\sigma n}$ are functions of the separation vector $\bm{r}'-\bm{r}$
that depend on the scattering length $a$ and the mass ratio $m_1/m_2$.
The coefficient of the leading operator is a Dirac delta function of  $\bm{r}'-\bm{r}$.
It can be derived simply by using the (anti)commutation relations 
for the field operators.  The coefficients of higher dimension operators
can be calculated using diagrammatic methods \cite{Braaten:2008uh,Braaten:2008bi}.

Upon inserting the OPE in Eq.~(\ref{OPE}),
the loss rate in Eq.~(\ref{dNdt-simple}) reduces to
%-----------------
\begin{eqnarray}
\frac{d \ }{dt} N_\sigma &=& 
- \frac{\hbar \, {\rm Im}(1/a)}{4 \pi \mu} 
\Bigg( \int \!\! d^3 R  \,
\big\langle |g_0|^2 \psi_1^\dagger \psi_2^\dagger \psi_2 \psi_1 (\bm{R}) \big\rangle 
+ \sum_n \int \!\! d^3 r \, f_{\sigma n}(\bm{r}) \,
 \int \!\! d^3 R\big\langle {\cal O}_n (\bm{R}) \big\rangle \Bigg) .
 \nonumber\\
\label{dNdt-normal}
\end{eqnarray}
%-----------------
Since the operators are all local, the restriction to the connected part of the 
matrix element is no longer necessary.
The expectation value in the first integral on the right side
can be identified as the contact density:
%-----------------
\begin{equation}
{\cal C}(\bm{R}) = 
\big\langle |g_0|^2 \psi_1^\dagger \psi_2^\dagger 
                    \psi_2 \psi_1 (\bm{R}) \big\rangle .
\label{condens-matel}
\end{equation}
%-----------------
This agrees with the expression for the contact density derived in 
Ref~\cite{Braaten:2008uh}, where $g_0$ was real.
The expression for the loss rate in Eq.~(\ref{dNdt-normal}) reduces to
%-----------------
\begin{eqnarray}
\frac{d \ }{dt} N_\sigma = 
- \frac{\hbar \, {\rm Im}(1/a)}{4 \pi \mu} 
\Bigg( C
+ \sum_n F_{\sigma n} \,
 \int \!\! d^3 R \big\langle {\cal O}_n (\bm{R}) \big\rangle \Bigg) ,
 \label{dNdt-int}
\end{eqnarray}
%-----------------
where $C$ is the contact.
The additional terms involve integrals over space of expectation values 
of the local operators ${\cal O}_n$.  Their coefficients 
$F_{\sigma n} = \int \!\!d^3 r \, f_{\sigma n}(\bm{r})$
are functions of $a$ and the mass ratio.
This is the general form of the universal relation for the 
inelastic two-atom loss rate in Eq.~(\ref{lossrate}).

The OPE in Eq.~(\ref{OPE}) is a systematic 
expansion in local operators with increasingly higher scaling dimension.
The scaling dimension is a property of an operator that determines the 
 behavior of its correlation functions at short distances.
The correlation function
$\langle {\cal O}(\bm{r})\, {\cal O}'(\bm{r}')\rangle $ for a pair of local operators 
generally diverges as a power of the separation $|\bm{r} - \bm{r}'|$
as the separation vector approaches 0.
The power is determined by the scaling dimensions of the operators. 
The quantum field operators $\psi_\sigma$ and the number density operators
$\psi_\sigma^\dagger \psi_\sigma$ have the naive scaling dimensions 
$\frac32$ and 3 of a nonrelativistic free field theory.
A gradient in an operator increases its scaling dimension by 1.
For example, the kinetic energy operators 
$\bm{\nabla}\psi_\sigma^\dagger \cdot \bm{\nabla}\psi_\sigma$
have  scaling dimension 5.  A complete set of dimension-5 operators 
is listed in Ref.~\cite{Nishida:2011xb}.
In a free field theory, the scaling dimension of a composite operator 
is the sum of the scaling dimensions of its factors.  Thus the naive scaling 
dimension of $\psi_1^\dagger \psi_2^\dagger \psi_2 \psi_1$ is 6.
However scaling dimensions can be strongly modified by interactions.
In the case of a large scattering length, the scaling 
dimension of $\psi_1^\dagger \psi_2^\dagger \psi_2 \psi_1$
is reduced from 6 to 4 \cite{Nishida:2007pj}.
Three-body operators, which contain 3 quantum fields and their Hermitian conjugates,
have non-integer scaling dimensions \cite{Nishida:2007pj}.  
In the case of fermions with equal masses ($m_1=m_2$), the lowest-dimension 
3-body operators that can appear in the OPE in Eq.~(\ref{OPE}) are
$\bm{{\cal O}}_\sigma^\dagger \cdot \bm{{\cal O}}_\sigma$, where 
$\bm{{\cal O}}_\sigma = 2 \psi_2 \psi_1 \bm{\nabla}\psi_\sigma 
- \psi_\sigma \bm{\nabla} ( \psi_2 \psi_1)$ and $\sigma$ is 1 or 2.
They have scaling dimension 8.54 
and the next higher operator has scaling dimension 9.33 \cite{Nishida:2011xb}.
In the case of unequal masses $m_1 < m_2$, the operator analogous to 
$\bm{{\cal O}}_1^\dagger \cdot \bm{{\cal O}}_1$ has a scaling dimension that 
decreases as $m_2/m_1$ increases.  It reaches 5 at the critical value 
$m_2/m_1 =13.6$ and remains at 5 for larger values of 
$m_2/m_1$ \cite{Nishida:2011xb}.
If atom 1 is a boson, the lowest-dimension 
3-body operator that can appear in the OPE in Eq.~(\ref{OPE}) is
$\psi_1^\dagger \psi_1^\dagger \psi_2^\dagger  \psi_2 \psi_1 \psi_1$.
The scaling dimensions of this operator is 5 for all values of the
mass ratio $m_2/m_1$.

We now consider which higher-dimension operators appear in the OPE for 
the loss rate in Eq.~(\ref{dNdt-normal}).  The only 
operators that can contribute are those for which the integral over all space 
of the expectation value $ \langle {\cal O}_n (\bm{R}) \rangle$ is nonzero.  
Operators that are total derivatives have expectation values whose integrals are zero.
If the system is homogeneous, only scalar operators have nonzero expectation values.  
If the system is static, operators that are total time derivatives have 
expectation values that are zero.
Linear combinations of operators that vanish upon using the equations of motion for the 
quantum field operators also have zero expectation values.
The additional operators that appear in the OPE in Eq.~(\ref{dNdt-normal})
are also limited to 3-body operators. 
This can be seen by using the (anti)commutation relations to normal order the 
quantum fields in the operator product in Eq.~(\ref{OPE}).
The result is the sum of the first term on the right side 
and the normal-ordered form of the left side.
But the normal-ordered operator has nonzero expectation value 
only in states that include 3 or more particles.
Thus only 3-body  and higher-body operators can appear in the additional terms 
in the loss rate in Eq.~(\ref{dNdt-normal}).
The contact operator in Eq.~(\ref{dNdt-normal}) has scaling dimension 4.
In the case of equal-mass fermions, the next term has an operator 
with scaling dimension 8.54.
In the case of fermions with $m_2/m_1 >13.6$ and in the case of bosons,
the next term has an operator with scaling dimension 5.

We can use dimensional analysis to determine how the coefficients  
$F_{\sigma n}$ of the higher-dimension terms in the loss rate  
in Eq.~(\ref{dNdt-int}) depend on $a$.  
The expectation value of an operator $ {\cal O}_n$ with scaling 
dimension $\Delta_n$ has dimensions (Length)$^{-\Delta_n}$.  
For example, the expectation value of the contact density operator
$ |g_0|^2 \psi_1^\dagger \psi_2^\dagger \psi_2 \psi_1$
has dimensions (Length)$^{-4}$.
By dimensional analysis applied to the expectation value of the OPE in 
Eq.~(\ref{OPE}), $f_{\sigma n}(\bm{r})$ must have dimensions
(Length)$^{\Delta_n-7}$.  Its integral over space $F_{\sigma n}$
must therefore have dimensions (Length)$^{\Delta_n-4}$.  
Now the only dimensional variables that $f_{\sigma n}(\bm{r})$ 
depends upon are $a$ and the separation vector $\bm{r}$.
After integrating over $\bm{r}$, the only dimensional variable 
that remains is $a$.  Thus $F_{\sigma n}$ must be
$|a|^{\Delta_n-4}$ multiplied by a numerical constant 
that depends on the mass ratio $m_2/m_1$
and may depend on the sign of $a$.

%%%%%%%%%%%%%%%%%%%%%%%%%%%%%%%%%%%%%%%%%%%%%%%
\subsection{Truncation of the OPE}
\label{sec:converge}
%%%%%%%%%%%%%%%%%%%%%%%%%%%%%%%%%%%%%%%%%%%%%%%

The simple universal relation for the loss rate in Eq.~(\ref{lossrate})
can be obtained from the general universal relation in Eq.~(\ref{dNdt-int})
by omitting the contributions from higher dimension operators.
To make our discussion of the truncation of the operator product expansion 
more concrete, we consider 
a balanced homogeneous gas of atoms with number densities 
$n_1 = n_2$ at temperature $T$ and in volume $V$.
The Fermi wavenumber is $k_F = (6 \pi^2 n_1)^{1/3}$.
The integrals $\int d^3r$ on the right side of Eq.~(\ref{dNdt-normal})
are simply $V$.  The contact density is proportional to 
$a^2 k_F^6$ in the BCS limit $a \to 0^-$, 
$k_F^3/a$ in the BEC limit $a \to 0^+$, and 
$k_F^4$ in the unitary limit $a \to \pm \infty$ \cite{Braaten:2010if}.
The coefficient $F_{\sigma n}$ in Eq.~(\ref{dNdt-int})
is proportional to $|a|^{\Delta_n-4}$
and the expectation value $\langle {\cal O}_n \rangle$
is $k_F^{\Delta_n}$ multiplied by a function of $k_F a$.

In the BCS and BEC limits $a \to 0^\mp$,
the coefficient $F_{\sigma n}$ can be combined with factors of 
$k_F$ from $\langle {\cal O}_n \rangle$ to get the dimensionless factor 
$(k_F |a|)^{\Delta_n-4}$.
Thus the contributions from higher dimension
operators are parametrically suppressed by powers of $k_F |a|$.
In the case of equal-mass fermions, the contribution to the loss rate from the 
first higher-dimension operator is suppressed by a factor of $(k_F |a|)^{4.54}$.
This parametric suppression of higher dimension operators 
in the BCS and BEC limits
justifies the truncation of the OPE for the loss rate in 
Eq.~(\ref{dNdt-int}) to obtain the simple universal relation in  
Eq.~(\ref{lossrate}).

In the unitary limit $a \to \pm \infty$,
the dimensionless factors of $(k_F |a|)^{\Delta_n-4}$
associated with the coefficients in the OPE diverge.
This suggests that the OPE for the loss rate breaks down in
the unitary limit.  However this conclusion is incorrect.
The divergence of the coefficient is a signal that 
the operator ${\cal O}_n$ has a normalization that is
inconvenient in the unitary limit.
The OPE in Eq.~(\ref{OPE}) must be valid in the unitary limit
for a suitable basis of local operators.
We have chosen operators ${\cal O}_n$ that are normalized
in such a way that the coefficients $F_{\sigma n}$ in Eq.~(\ref{dNdt-int})
are proportional to $|a|^{\Delta_n-4}$.
As $ a \to \pm \infty$, the matrix element $\langle {\cal O}_n \rangle$
must therefore include a multiplicative factor of $|a|^{4-\Delta_n}$
that cancels the dependence of the coefficient on $a$.
To obtain an operator whose expectation value has a smooth unitary limit,
we can multiply the operator ${\cal O}_n(\bm{r})$
by $(\Lambda |a|)^{\Delta_n-4}$, where $\Lambda$ is an arbitrary 
momentum scale.  The expectation value of the resulting operator 
in the homogeneous Fermi gas in the unitary limit is proportional 
to $\Lambda^{\Delta_n-4} k_F^4$.  The dependence on $\Lambda$
is cancelled by a factor of  $\Lambda^{4-\Delta_n}$
in $F_{\sigma n}$, so the contribution to the loss rate is
proportional to $k_F^4$.
Thus all the terms inside the parentheses in Eq.~(\ref{dNdt-int})
will have the same parametric behavior $V k_F^4$ as the contact term. 
This suggests that all the terms are equally important.
If this was the case, one could not truncate the expansion
to obtain the simple universal relation in  Eq.~(\ref{lossrate}).

There is reason to believe that the simple universal relation
for the loss rate in Eq.~(\ref{lossrate}) remains a good approximation 
even near the unitary limit $a \to \pm \infty$.  There is no parametric 
suppression of higher dimension operators in this limit,
so the validity of the truncation depends on the convergence properties of the OPE.
The OPE was discovered independently by Kadanoff, Polyakov, and Wilson
in 1969 \cite{Kadanoff:1969,Polyakov:1969,Wilson:1969}.
Wilson proposed that the OPE was an asymptotic expansion 
in the small separation $| \bm{r} - \bm{r}' |$ of the operators \cite{Wilson:1969}.
Studies of the convergence of the OPE in relativistic quantum field theories
suggest that it is actually a convergent expansion.
Mack showed in 1977 that the OPE for a conformally-invariant field theory
in 4 space-time dimensions converges  for finite $| \bm{r} - \bm{r}' |$,
at least when acting on the vacuum state \cite{Mack:1976pa}.
More recently, Pappadopulo, Rychkov, Espin, and Rattazzi showed that
the OPE for a conformal field theory
in any space-time dimension not only has infinite radius of convergence in the separation
$| \bm{r} - \bm{r}' |$, but at fixed $| \bm{r} - \bm{r}' |$ 
the expansion converges absolutely with a convergence rate that is
exponential in the scaling dimension
\cite{Pappadopulo:2012jk}.  If the OPE is truncated by omitting terms with
scaling dimension $\Delta_0$ and higher, the error decreases
like $\exp(- A \Delta_0)$, where $A$ depends on the separation.
Furthermore, Hollands and Kopper showed that 
the OPE for the simplest interacting scalar field theory
in 4 Euclidean space dimensions has similar convergence properties
order by order in perturbation theory \cite{Hollands:2011gf}.
It has infinite radius of convergence in the separation
$| \bm{r} - \bm{r}' |$, and the convergence at fixed $| \bm{r} - \bm{r}' |$ 
is exponential in the scaling dimension.
If the OPE is truncated  by omitting terms with
scaling dimension $\Delta_0$ and higher, the error decreases
like $A ^{\Delta_0} /  (\Delta_0!)^{1/2}$, 
where $A$ depends on the state and is proportional to the separation \cite{Hollands:2011gf}.

The convergence of the OPE in relativistic field theories
does not guarantee its convergence in nonrelativistic field theories.
However it is at least plausible that the OPE in a nonrelativistic field theory
has infinite radius of convergence in the separation
$| \bm{r} - \bm{r}' |$, and that at fixed $| \bm{r} - \bm{r}' |$ 
the expansion converges absolutely with a convergence rate that is
 exponential in the scaling dimension.
 We will assume this is the case.
The OPE for the loss rate
in Eq.~(\ref{dNdt-normal}) then has exponential convergence 
in the scaling dimension.  In the unitary limit,
the higher dimension contributions have no parametric suppression,
but they do have numerical suppression.
In the case of a balanced gas of 
equal-mass fermions at zero temperature near the unitary limit, 
all the terms have the same parametric behavior $k_F^4$ 
as the contact density term, but the higher dimension terms 
are suppressed by numerical factors that decrease exponentially 
fast with the scaling dimension.
The large gap between the dimension 4 of the contact density operator 
and the dimension 8.54 of the next higher dimension operator
suggests that the the higher dimension terms in the loss rate 
may be very small.
Thus the simple universal relation
in Eq.~(\ref{lossrate}) obtained by truncating the OPE after the leading term
may be a good approximation even in the unitary limit.

%%%%%%%%%%%%%%%%%%%%%%%%%%%%%%%%%%%%%%%%%%%%%%%%
\section{Contact in Thermal Equilibrium}
\label{sec:ConPair}
%%%%%%%%%%%%%%%%%%%%%%%%%%%%%%%%%%%%%%%%%%%%%%%%

In this section, we calculate the contact for a single pair of atoms 
in thermal equilibrium to leading order in the imaginary part of the scattering length.
We then use the result to obtain the contact density for a dilute gas
of atoms or dimers.

\subsection{Definition of the contact}

In the absence of 2-atom inelastic scattering channels,
the scattering length $a$ is real.
In this case, the adiabatic relation in Eq.~(\ref{adiabatic-Tan}) can be used as an
operational definition of the contact. It can be generalized 
to nonzero temperature by taking the derivative with the entropy fixed.
Equivalently, the adiabatic relation can be expressed in the form
%-----------------
\begin{equation}
\frac{d \ \ }{da^{-1}} F = - \frac{\hbar^2}{8\pi \mu} ~ C ,
\label{adiabatic-F}
\end{equation}
%-----------------
where $F$ is the free energy and the derivative is taken with the temperature fixed.  
The free energy can be determined from the partition function:
%-----------------
\begin{equation}
Z = \exp(- \beta F) ,
\label{Z-F}
\end{equation}
%-----------------
where $\beta = 1/(k_B T)$.
One advantage of using the adiabatic relation as a definition of the contact
is that it is applicable in any formalism.

If there are inelastic scattering channels, the scattering length is complex.
In this case, Eq.~(\ref{condens-matel})
provides a field-theoretic definition of the contact density as the expectation value 
of an operator constructed out of quantum fields.
It would be useful to have a more general definition for complex $a$
that does not depend on any specific formalism.  
In particular, it would be good to have a 
generalization of the adiabatic relation in Eq.~(\ref{adiabatic-F}).
We have not constructed such a generalization.
Instead we will exploit the fact that the adiabatic relation in Eq.~(\ref{adiabatic-F}) 
provides a definition of the contact 
that is correct to leading order in the imaginary part of $1/a$.
In the remainder of this section, we ignore the effects of inelastic scattering channels
and take the scattering length $a$ to be real.

\subsection{Single pair of atoms}

We consider two atoms that are restricted to a volume $V$ 
that is much larger than $|a|^3$ and are in thermal equilibrium at temperature $T$.
We will use the universal relation in Eq.~(\ref{adiabatic-F}) 
to calculate the contact for the single pair of atoms.
In the center-of-mass frame, the spectrum of the pair includes 
a continuum above the threshold from scattering states.
Their energies can be parameterized 
as $E = \hbar^2 k^2/(2 \mu)$, where $k$ is the wavenumber.
The phase shift for an S-wave scattering state is
%-----------------
\begin{equation}
\delta(k) = - \arctan(k a) .
\label{delta-S}
\end{equation}
%-----------------
If $a>0$, the spectrum also has a discrete energy below the threshold:
the dimer, whose energy is $E = -E_D$, where $E_D$ is the binding energy:
%-----------------
\begin{equation}
E_D = \frac{\hbar^2}{2 \mu a^2} .
\label{E-dimer}
\end{equation}
%-----------------

Since the center-of-mass motion and the relative motion of the pair decouple,
the partition function is the product of the associated partition functions
$Z_{\rm cm}$ and $Z_{\rm rel}$.  The dependence on $a$ enters only through 
$Z_{\rm rel}$.  In the limit $a \to 0^-$, there are no interactions and 
$Z_{\rm rel}$ reduces to the partition function for a free particle.
%-----------------
\begin{eqnarray}
Z_{\rm free} = V \int \frac{d^3k}{(2 \pi)^3}
e^{- \beta \hbar^2 k^2/(2 \mu)} 
= V /\lambda_T^3,
\label{Z-free}
\end{eqnarray}
%-----------------
where $V$ is the spatial volume and $\lambda_T$ is the thermal wavelength
for a particle of mass $\mu$:
%-----------------
\begin{equation}
\lambda_T = \sqrt{\frac{2 \pi \hbar^2}{\mu k_B T}}.
\label{lambda-T}
\end{equation}
%-----------------
We express $Z_{\rm rel}$ as the sum of the free partition function in
Eq.~(\ref{Z-free}) and an interaction term that has all the dependence 
on $a$: $Z_{\rm rel} = Z_{\rm free} + Z_{\rm int}$.
The interaction term receives contributions from the continuum
of S-wave scattering states and also, if $a > 0$, from the dimer. 
The spectrum of scattering states can be made discrete with energies
$E_n$ by making the spatial volume compact.  A convenient method 
for discretizing the scattering states is to put the system in a spherical 
box with a large radius $R$.  
The interaction term can then be expressed as
%-----------------
\begin{equation}
Z_{\rm int}(a) = \theta(a) e^{\beta E_D} 
+ \sum_{n=1}^\infty e^{- \beta \hbar^2 k_n^2/(2 \mu)}
- \sum_{n=1}^\infty e^{- \beta \hbar^2 k_n^2/(2 \mu)} \Big|_{a \to 0^-} .
\label{Z-int}
\end{equation}
%-----------------
The scattering wavefunctions
$\sin[k r + \delta(k)]$ must satisfy the boundary condition
%-----------------
\begin{equation}
\sin[k_n R + \delta(k_n)] = 0.
\label{sin-0}
\end{equation}
%-----------------
Its nontrivial solutions satisfy
%-----------------
\begin{equation}
k_n R - \arctan( k_n a)  = n \pi, ~~~~~~ n=1,2,3,... \,.
\label{k-n}
\end{equation}
%-----------------
In the limit $R \to \infty$, the discrete energies approach a continuum 
and the sum over integers approaches an integral:
%-----------------
\begin{equation}
\sum_{n=1}^\infty \longrightarrow 
\frac{1}{\pi} \int_0^\infty dk \, 
\left( R - \frac{a}{1 + k^2 a^2} \right) .
\label{sum-int}
\end{equation}
%-----------------
Inserting this into Eq.~(\ref{Z-int}), we find that the dependence on $R$ 
cancels and the interaction term reduces to
%-----------------
\begin{equation}
Z_{\rm int} = \theta(a) e^{\beta \hbar^2 /(2 \mu a^2)} 
- \frac{a}{\pi} \int_0^\infty dk \, \frac{1}{1 + k^2 a^2}
e^{- \beta \hbar^2 k^2/(2 \mu)}.
\label{Z-int:a}
\end{equation}
%-----------------

The universal relation in Eq.~(\ref{adiabatic-F}) implies that 
the contact for the single pair of atoms is given by
%-----------------
\begin{equation}
C_{\rm pair} = \frac{8 \pi \mu}{\hbar^2 \beta} 
\frac{d \ \ }{da^{-1}} \log (Z_{\rm free} + Z_{\rm int}).
\label{contact-Z}
\end{equation}
%-----------------
The derivative of  $Z_{\rm int}$ is evaluated most easily if we first change the integration
variable in Eq.~(\ref{Z-int:a}) to the dimensionless variable $x = a k$.
The resulting expression  is
%-----------------
\begin{equation}
\frac{8 \pi \mu}{\hbar^2 \beta} \frac{d \ \ }{da^{-1}} Z_{\rm int} =  
\theta(a) \, \frac{8 \pi}{a} e^{\beta \hbar^2 /(2 \mu a^2)} 
+ 8 a^2 \int_0^\infty dk \, 
\frac{k^2}{1 + a^2 k^2} e^{- \beta \hbar^2 k^2/(2 \mu)} .
\label{dZint}
\end{equation}
%-----------------
Our final result for the contact for a single pair of atoms  is
%-----------------
\begin{equation}
C_{\rm pair} = \frac{8 \pi \mu}{\hbar^2 \beta}
\frac{(d /da^{-1}) Z_{\rm int}}
{V/ \lambda_T^3 +  Z_{\rm int}} ,
\label{contact-Z:V}
\end{equation}
%-----------------
where $Z_{\rm int}$ and its derivative are given in 
Eqs.~(\ref{Z-int:a}) and (\ref{dZint}).
The integrals in these expressions can be evaluated analytically in
terms of error functions.

The contact in Eq.~(\ref{contact-Z:V}) is for a pair of atoms in thermal equilibrium 
and also in chemical equilibrium with respect to transitions between a dimer 
and two unbound atoms.
The contact for two unbound atoms that are not in 
chemical equilibrium with the dimer can be obtained by omitting the
dimer contributions in the numerator and denominator of Eq.~(\ref{contact-Z:V}).
In the large volume limit, we need only keep the first term in the denominator. 
The contact for the two unbound atoms reduces to
%-----------------
\begin{equation}
C_{AA} =  
\frac{8 a^2 \lambda_T^3}{V}
\int_0^\infty dk \, 
\frac{k^2}{1 + a^2 k^2} e^{- \beta \hbar^2 k^2/(2 \mu)} .
\label{C-2atoms}
\end{equation}
%-----------------
The contact for a dimer that is not in chemical equilibrium with unbound atoms
can be obtained by keeping only the dimer terms in the numerator and 
denominator of Eq.~(\ref{contact-Z:V}).  The result is extremely simple:
%-----------------
\begin{equation}
C_D =  
\frac{8 \pi}{a}.
\label{C-dimer}
\end{equation}
%-----------------

\subsection{Dilute gas of atoms or dimers}

We can use the expression in Eq.~(\ref{C-2atoms}) for the contact 
for a single pair of unbound atoms to deduce the contact density for a 
homogeneous gas of atoms in the dilute limit.
If the numbers of atoms in the volume $V$ are $N_1$ and $N_2$,
the total number of pairs is $N_1 N_2$.
Each pair gives the same contribution $C_{AA}$ to the contact.
We obtain the contact density by dividing by the volume:
${\cal C}_{AA} = N_1 N_2 C_{AA}/V$.  Using the expression for $C_{AA}$ in
Eq.~(\ref{C-2atoms}), we find that the contact density is
proportional to the product of the number densities 
$n_1$ and $n_2$:
%-----------------
\begin{equation}
{\cal C }_{AA}  = 8 a^2 \lambda_T^3
\left( \int_0^\infty \!\!dk\,\frac{k^2}{1 + a^2 k^2} e^{- \beta \hbar^2 k^2/(2 \mu)} \right)
n_1 n_2.
\label{contactdensity}
\end{equation}
%-----------------

We can use the expression in Eq.~(\ref{C-dimer}) for the contact $C_D$
of a dimer to deduce the contact density for a 
homogeneous gas of dimers in the dilute limit.
The total contact is $C_D$ multiplied by the number $N_D$ of dimers.
We obtain the contact density by dividing by the volume.
The result is proportional to the number density
$n_D$ of the dimers:
%-----------------
\begin{equation}
{\cal C}_D =  
\frac{8 \pi}{a} n_D.
\label{C-dimergas}
\end{equation}
%-----------------

%%%%%%%%%%%%%%%%%%%%%%%%%%%%%%%%%%%%%%%%%%%%%%%%
\section{Direct Calculations of Inelastic Loss Rates}
\label{sec:DirLoss}
%%%%%%%%%%%%%%%%%%%%%%%%%%%%%%%%%%%%%%%%%%%%%%%%

In this section, we carry out direct calculations of inelastic loss rates 
for dilute homogeneous gases of atoms or dimers in thermal equilibrium.
We take the scattering length $a$ to be complex, with a negative 
imaginary part that
takes into account inelastic scattering channels.
For the direct calculations, it is more convenient to use the
inverse scattering length $\gamma = 1/a$, which 
has a positive imaginary part.

\subsection{Low-density gas of atoms}

The elastic scattering amplitude for atoms 
with a large complex scattering length $a=1/\gamma$ at wavenumber $k$ is
%----------------------
\begin{eqnarray}
f(k) = \frac{1}{-\gamma-ik}\,.
\label{scattamp}
\end{eqnarray}
%----------------------
The differential cross section $|f(k)|^2$ is isotropic.
The elastic cross section is obtained by multiplying it by the solid angle:
%----------------------
\begin{eqnarray}
\sigma^{\rm (elastic)}(k) = \frac{4 \pi}{\left| -\gamma-ik\right|^2}\,.
\label{elacross}
\end{eqnarray}
%----------------------
According to the optical theorem, the total cross section is the
imaginary part of the forward scattering amplitude multiplied by $4\pi/k$:
%----------------------
\begin{eqnarray}
\sigma^{\rm (total)}(k) 
= \frac{4 \pi}{k}\Im\left(\frac{1}{-\gamma-ik}\right)\,.
\label{totcross}
\end{eqnarray}
%----------------------
The inelastic cross section  is
the difference between the total and elastic cross sections
in Eqs.~(\ref{totcross}) and (\ref{elacross}).
The inelastic transition rate $g(k)$ per volume squared can be obtained
by dividing the inelastic cross section by the flux factor $\mu/\hbar k$:
%----------------------
\begin{eqnarray}
g(k)&=&\frac{4\pi\hbar}{\mu}\,\frac{\Im(\gamma)}{\left| -\gamma-ik\right|^2}\,.
\label{gf2}
\end{eqnarray}
%----------------------

In the dilute limit, the inelastic loss rate can be written
%----------------------
\begin{eqnarray}
\frac{d}{dt}n_1=\frac{d}{dt}n_2= - K_2 \,n_1 n_2\,,
\label{betadef}
\end{eqnarray}
%---------------------- $
where $n_1$ and $n_2$ denote the number densities of atoms 1
and 2, respectively.
For an ensemble of particles 1 and 2, the inelastic loss rate coefficient $K_2$
can be calculated via a statistical average of the inelastic transition rate $g(k)$. 
Taking the ensemble to be atoms 
in thermal equilibrium at temperature $T$, the rate coefficient is
%----------------------
\begin{eqnarray}
K_2 &=& \frac{\int d^3 k \,e^{-\beta\hbar^2 k^2/(2\mu)}\, g(k)}
{\int d^3 k\,  e^{-\beta\hbar^2 k^2/(2\mu)}}
= \frac{\lambda_T^{3}}{2\pi^2} \int_0^\infty dk \,k^2 \,
e^{-\beta\hbar^2 k^2/(2\mu)}\, g(k)\,.
\label{betaT0}
\end{eqnarray}
%----------------------
Inserting the inelastic transition rate $g(k)$ in Eq.~(\ref{gf2}), we obtain
the loss rate coefficient $K_2$:
%----------------------
\begin{eqnarray}
K_2 = \frac{2\hbar \lambda_T^{3}}{\pi\mu}\, {\rm Im}(\gamma)
\int_0^\infty 
dk \,\frac{k^2}{| -\gamma-ik |^2} e^{-\beta\hbar^2 k^2/(2\mu)}\,.
\label{betaT1}
\end{eqnarray}
%----------------------

According to the universal relation in Eq.~(\ref{lossrate}),
the right side of Eq.~(\ref{betadef}) must be proportional to the contact density:
%-----------------
\begin{equation}
K_2 n_1 n_2 = \frac{\hbar}{4 \pi \mu}\, {\rm Im}(1/a
)\, {\cal C}_{AA} .
\label{K2-C}
\end{equation}
%-----------------
Comparing the expression for  $K_2$ in Eq.~(\ref{betaT1}) and ${\cal C}_{AA}$ 
in  Eq.~(\ref{contactdensity}), we see that they agree only to leading order in 
Im($a$).  This is not surprising, because the contact density in
Eq.~(\ref{contactdensity}) was calculated 
only for the case of real $a$.  That calculation could presumably be extended
to the case of complex $a$.  The result must agree with Eq.~(\ref{K2-C})
in the dilute limit.  Thus the contact density in the dilute limit for 
atoms with a complex scattering length $a$ must be
%----------------------
\begin{eqnarray}
{\cal C}_{AA} = 8 \lambda_T^{3}
\left( \int_0^\infty 
dk \,\frac{k^2 e^{-\beta\hbar^2 k^2/(2\mu)}}{| -1/a-ik |^2}
\right) n_1 n_2 \,.
\label{Ccomplexa}
\end{eqnarray}
%----------------------

\subsection{Low-density gas of dimers}

The inelastic loss rate for a dilute, noninteracting gas of dimers
can be written as
%-----------------
\begin{equation}
\frac{d\ }{dt} n_D = - \Gamma_D  n_D,
\label{dnD-tau}
\end{equation}
%-----------------
where $\Gamma_D$ is the decay rate or inverse lifetime of the dimer.
The decay rate can be expressed as an integral over momenta of the
dimer wave function $\psi(k)$ in momentum space, its complex conjugate  $\psi^*(k')$,
and the imaginary part of the two-particle-irreducible transition 
amplitude $U(k,k')$ for a pair of atoms:
%-----------------
\begin{equation}
\Gamma_D = \frac{2}{\hbar} \int \frac{d^3 k}{(2\pi)^3} 
\int \frac{d^3 k}{(2\pi)^3}\psi(k) \left( - {\rm Im}\,U(k,k') \right) \psi^*(k').
\label{eq:GammaD}
\end{equation}
%-----------------
In the field theory formulation of the problem with the interaction term in
Eq.~(\ref{U}), $U(k,k')$ is simply the bare coupling constant $g_0 \,\hbar^2
/(2\mu)$. Since this is momentum independent, $\Gamma_D$ simplifies to
%-----------------
\begin{equation}
\Gamma_D =  \frac{\hbar}{\mu} \left( -{\rm Im}\, g_0 \right)
\left| \int^{(\Lambda)}  \!\!\! \frac{d^3 k}{(2\pi)^3}\, \psi(k)\right|^2.
\label{eq:GammaD-int}
\end{equation}
%-----------------
The normalized dimer wave function for complex scattering 
length $a=1/\gamma$ is
%-----------------
\begin{equation}
\psi(k)=\frac{\sqrt{8\pi{\rm Re}(\gamma)}}{\gamma^2+k^2}.
\end{equation}
%-----------------
If the integral in Eq.~(\ref{eq:GammaD-int}) was convergent,
it would be the wave function at the origin in position space.
However it is ultraviolet divergent, and it has therefore been regularized 
with a momentum cutoff $\Lambda$. 
The regularized integral is
%-----------------
\begin{equation}
\int_0^\Lambda \frac{k^2 dk}{2\pi^2} \,\psi(k)=
\frac{\sqrt{8\pi{\rm Re}(\gamma)}}{2\pi^2}\left(\Lambda-\gamma\pi/2\right).
\label{eq:intpsi}
\end{equation}
%-----------------
This dependence on $\Lambda$ is precisely what is required to cancel
the $\Lambda$-dependence of the factor of Im$\,g_0$ in Eq.~(\ref{eq:GammaD-int}).
The imaginary part of the coupling constant $g_0$ in Eq.~(\ref{g})
can be expressed as
%-----------------
\begin{equation}
{\rm Im}\,g_0=-\frac{\pi^3 \,{\rm Im}(\gamma)}{\left|\Lambda-
\gamma\pi/2\right|^2}.
\label{eq:Img0}
\end{equation}
%-----------------
Inserting Eqs.~(\ref{eq:Img0}) and (\ref{eq:intpsi}) into Eq.~(\ref{eq:GammaD-int}),
we obtain a simple result for the decay rate:
%-----------------
\begin{equation}
\Gamma_D = \frac{2\hbar}{\mu}{\rm Re}(\gamma)\,{\rm Im}(\gamma)
= \frac{2\hbar \,{\rm Re}(a)\,{\rm Im}\,(1/a)}{\mu |a|^2}.
\label{GamD_exact}
\end{equation}
%-----------------

This result for $\Gamma_D$ can also be obtained from the expression for the
dimer binding energy in Eq.~(\ref{E-dimer}) by analytically continuing the
scattering length $a$ to complex values.
By expressing the complex energy $-\hbar^2/(2\mu a^2)$
in the form $-E_D -i\hbar\Gamma_D/2$, we can read off the width 
$\hbar\Gamma_D$ of the resonance.
This prescription gives the same result for 
$\Gamma_D$ as Eq.~(\ref{GamD_exact}).

According to the universal relation in Eq.~(\ref{lossrate}),
the right side of Eq.~(\ref{dnD-tau}) must be proportional to the contact density:
%-----------------
\begin{equation}
\Gamma_D n_D = \frac{\hbar}{4 \pi \mu}\, {\rm Im}(1/a)\, {\cal C}_D .
\label{K2-CD}
\end{equation}
%-----------------
Comparing the expression for  $\Gamma_D$ in Eq.~(\ref{GamD_exact})
and ${\cal C}_D$ in  Eq.~(\ref{C-dimergas}), we see that they agree only 
to next-to-leading order in Im($a$).  
This is not surprising, because the contact density 
in Eq.~(\ref{C-dimergas}) was calculated 
only for the case of real $a$.  That calculation could presumably be extended
to the case of complex $a$.  The result must agree with Eq.~(\ref{K2-CD})
in the dilute limit.  Thus the contact density for a dilute gas of dimers
in the case of a complex scattering length $a$ must be
%-----------------
\begin{equation}
{\cal C}_D =  
\frac{8 \pi\,{\rm Re}(a)}{|a|^2} n_D.
\label{C-dimercomplexa}
\end{equation}
%-----------------

K\"ohler, Tiesinga, and Julienne
have derived a universal relation between the inelastic rate coefficient $K_2$ 
and the dimer lifetime $1/\Gamma_D$
\cite{KTJ08}:\footnote{This is the relation for fermions with two spin states. 
For identical bosons, the proportionality factor is $4\pi a^3$~\cite{KTJ08}.}
%-----------------
\begin{equation}
K_2 = 2 \pi a^3 \Gamma_D .
\label{dnD-tau2}
\end{equation}
%-----------------
They considered the case in which Im$(a)$ is negligible compared to Re$(a)$.
Using the expressions for $K_2$ in Eq.~(\ref{betaT1}) and $\Gamma_D$ in Eq.~(\ref{GamD_exact}), we can see that the relation holds for $K_2$ 
in the zero-temperature limit and we can generalize it to a complex scattering length $a$:
%-----------------
\begin{equation}
K_2(T=0) = \frac{2 \pi |a|^4}{{\rm Re}(a)} \Gamma_D.
\label{dnD-tau2c}
\end{equation}
%-----------------
This differs from the relation for real $a$ in Eq.~(\ref{dnD-tau2}) 
at next-to-next-to-leading order in ${\rm Im}(a)$.

%%%%%%%%%%%%%%%%%%%%%%%%%%%%%%%%%%%%%%%%%%%%%%%
\section{Trapped gas of equal-mass fermions}
\label{sec:applications}
%%%%%%%%%%%%%%%%%%%%%%%%%%%%%%%%%%%%%%%%%%%%%%%

The universal relation for the inelastic two-body loss rate in Eq.~(\ref{dNdt-normal})
can be tested experimentally using atoms in hyperfine states that have
an inelastic spin relaxation channel.  Such an experiment has been carried out 
for identical bosons by Thompson, Hodby, and Wieman using an ultracold trapped
thermal gas of $^{85}$Rb atoms near a Feshbach resonance \cite{THE08}.
The results were interpreted theoretically by K\"ohler, Tiesinga, and Julienne
in terms of the spontaneous dissociation of dimers \cite{KTJ08}.
The calculated lifetime of the dimers agreed well with the results extracted from the 
experiment except very close to the Feshbach resonance, 
where the predicted lifetime was too large.
The extension of our universal relation for the atom loss rate 
to the case of identical bosons should allow a quantitative understanding of
the loss rate much closer to the Feshbach resonance.
Identical bosons introduce additional complications.
One theoretical complication is that the operator product expansion 
analogous to Eq.~(\ref{OPE}) includes the 3-body contact operator,
which has scaling dimension 5 \cite{Braaten:2011sz}.  
The contribution from the 3-body contact may therefore not be strongly suppressed 
compared to the leading term from the 2-body contact.
An experimental complication is that the contribution to the 
loss rate from 3-body recombination into the shallow dimer 
may also be important  \cite{THE08}.
It is possible that part of the loss rate attributed to  3-body recombination
in the experiment in Ref.~\cite{THE08} was actually due to the contact 
from unbound atoms.

An analogous experiment with equal-mass fermions would 
be particularly easy to interpret.  The severe suppression of higher dimension 
operators implies that the simple universal relation in Eq.~(\ref{lossrate})
should be very accurate.  Additional contributions to the 
loss rate from 3-body recombination into the shallow dimer 
are also suppressed.  The experiment could be carried out in the same way 
as the experiment with  $^{85}$Rb atoms in Ref.~\cite{THE08}.
It would begin with a large number of atoms in thermal equilibrium
at a scattering length on the negative side of a Feshbach resonance.
The scattering length would then be ramped to a large positive value.
The number of dimers that are produced could be controlled 
by varying the ramp rate. The scattering length would be held at the 
large positive value for a variable time, during which atoms and dimers 
could disappear.  Finally, the scattering length would 
be ramped back to the negative side of the Feshbach resonance to dissociate
the dimers, and the total number of remaining atoms would then be measured.
By analyzing the dependence of the number of remaining atoms on the holding time,
one could determine the contact for an unbound atom pair and/or the contact for a dimer.
These contacts could be measured as functions of the scattering length
and compared with the universal predictions.

We therefore consider a dilute gas of fermionic atoms 
and dimers that are trapped in a harmonic potential
and in thermal equilibrium at a very low temperature $T$.
We take the two types of atoms to be different spin states, 
so they have equal masses $m_1 = m_2 = m$, and we assume that
they both feel the same trapping potential:
%-----------------
\begin{equation}
V(\bm{r}) = \frac12 m (\omega_x^2 x^2  + \omega_y^2 y^2 + \omega_z^2 z^2) .
\label{Vtrap}
\end{equation}
%-----------------
The number densities $n_1$ and $n_2$ for the two spin states
are Gaussians whose integrals over space are the total numbers of 
atoms $N_1$ and $N_2$, respectively.  The dimers feel the potential 
$2 V(\bm{r})$.  Their number density $n_D$ is also a
Gaussian whose integral over space is the total number of 
dimers $N_D$. 

For simplicity, we assume that the imaginary part of the scattering length
is extremely small compared to its real part.
It therefore enters only in the prefactor of the contact in the  
universal relation in Eq.~(\ref{dNdt-normal}).
The contact has contributions from unbound atoms and from dimers.
The contribution from unbound atoms can be obtained
by using the result for the contact density in Eq.~(\ref{contactdensity})
together with the local density approximation.
It is obtained 
by integrating the contact density in Eq.~(\ref{contactdensity}) over all space.
The result can be expressed as $N_1 N_2 C_{AA}$, 
where the contact for a single pair 
of atoms is
%-----------------
\begin{equation}
C_{AA} = 8 a^2
\left( \frac{\hbar \bar \omega}{k_B T} \right)^3
\int_0^\infty \!\!dk\,\frac{k^2}{1 + a^2 k^2} e^{- \beta \hbar^2 k^2/m} 
\label{contacttrap}
\end{equation}
%-----------------
and $\bar \omega = (\omega_x  \omega_y  \omega_z)^{1/3}$
is the geometric mean of the trapping frequencies.
The contribution to the contact from dimers can be obtained
by using the result for the contact density in Eq.~(\ref{C-dimergas})
together with the  local density approximation.
Integrating the contact density over all space, we obtain the simple result
$N_D C_D$, where $C_D= 8 \pi a/m$ is the contact for a single dimer.

The number of dimers evolves independently from the number of unbound atoms.
The time dependence for the number of dimers is
%-----------------
\begin{equation}
N_D(t) = N_D(0) \exp(- \Gamma_D\, t),
\label{ND-t}
\end{equation}
%-----------------
where $\Gamma_D=4\hbar\,{\rm Im}\,(1/a)/(ma)$.
If the system is balanced, with equal numbers of atoms in the two spin states,
the time dependence for the number of unbound atoms is also simple:
%-----------------
\begin{equation}
N_\sigma(t) = 
\left[ \frac{1}{N_\sigma(0)} + \frac{\hbar\,  \Im(1/a)}{2 \pi m} C_{AA} \,t 
\right]^{-1},
\label{NA-t}
\end{equation}
%-----------------
where $C_{AA}$ is the contact for a pair of unbound atoms in Eq.~(\ref{contacttrap}).
If the initial number of dimers is sufficiently large,
the loss rate will at first be dominated by the dimer contact.
However, as the dimers decay away, the loss rate will eventually be dominated by
the contact from unbound atoms.  The cross-over point at which the dominant 
loss mechanism changes from the dimer contact to the contact from unbound atoms
is when the numbers of atoms 
and dimers satisfy $N_1 N_2/N_D \approx C_D/C_{AA}$.

%%%%%%%%%%%%%%%%%%%%%%%%%%%%%%%%%%%%%%%%%%%%%%%
\section{Summary}
\label{sec:summary}
%%%%%%%%%%%%%%%%%%%%%%%%%%%%%%%%%%%%%%%%%%%%%%%

In summary, we have developed a rigorous theoretical framework 
for calculating the inelastic two-body loss rate for ultracold atoms
with a large scattering length.
We used the operator product expansion (OPE) to derive the universal relation 
for the inelastic two-body loss rate.  The general result is given in
Eq.~(\ref{dNdt-normal})  as an expansion
in terms of integrals of expectation values
of local operators with increasingly higher scaling dimensions.
If we truncate the OPE to the leading term, we get the simple
universal relation in Eq.~(\ref{lossrate}) between the loss rate and the contact.
In the weak-coupling limit in which the 
largest  important momentum scale $k_0$ of the system satisfies $k_0 |a| \ll 1$,
the OPE provides a systematically improvable approximation to the loss rate,
because higher dimension terms are parametrically suppressed 
by powers of $k_0 |a|$. 
The truncation of the OPE to the leading term
is particularly accurate in the case of fermions with equal masses,
because the first correction is suppressed by a factor of $(k_0 |a|)^{4.54}$.
If the OPE has the same convergence properties as in relativistic field theories,
the general universal relation for the loss rate in Eq.~(\ref{dNdt-normal})  
may remain useful near the unitary limit.
If $k_0 |a| \gg 1$, the higher dimension terms have the same
parametric dependence on $k_0$ as the contact density term.
However the higher density terms may have a  numerical suppression
that is exponential in the scaling dimension.
This could justify the truncation of the OPE to the leading term,
so the simple universal relation in Eq.~(\ref{lossrate}) between the loss rate 
and the contact may be a good approximation even in the unitary limit.

We verified the universal relation by direct calculations for homogeneous gases 
of atoms and dimers in thermal equilibrium in the dilute limit.
The contact density for a dilute gas of atoms is given in
Eq.~(\ref{contactdensity}) for the case of a real scattering length  and in
Eq.~(\ref{Ccomplexa}) for the case of a complex scattering length.
The corresponding results for a dilute gas of dimers are given in
Eqs.~(\ref{C-dimergas}) and (\ref{C-dimercomplexa}).

We described how the universal relation  in Eq.~(\ref{lossrate})
could be tested using dilute ultracold gases of fermions with two spin states.
The time dependence of the loss rate at a fixed scattering length
could be used to measure both the contact for a single dimer 
and the contact for a pair of unbound atoms.
The dependence of these contacts on the scattering length could then be 
compared with universal predictions.
Agreement with the predictions would provide a beautiful 
illustration of how the universal relations constrain the interplay between 
few-body and many-body physics.

\begin{acknowledgments}
This research was supported in part by the Department of Energy 
under grant DE-FG02-05ER15715, by the Army Research Office,
by the National Science Foundation,
by the DFG and the NSFC through 
funds provided to the Sino-German CRC 110, by the BMBF under 
grant 05P12PDFTE, and by the Alexander von Humboldt Foundation.
\end{acknowledgments}

% \begin{appendix}
%%%%%%%%%%%%%%%%%%%%%%%%%%%%%%%%%%%%%%%%%%%%%%%%%%%%%%%%%%%%%%
%\section{Diagrammatic Calculations}
% \end{appendix}

%

\end{document}